\documentstyle[aps,epsf,preprint]{revtex}

\begin{document}

\title{Three-Point Spectral Functions in $\phi^3_6$ Theory at 
Finite Temperature}

\author{Hou Defu\cite{email} and U. Heinz}

\address{Institut f\"ur Theoretische Physik, Universit\"at Regensburg, \\
D-93040 Regensburg, Germany}

\date{\today}
\preprint{hep-th/9704392}

\maketitle

\begin{abstract}

We derive a set of relations among the thermal components of the 
3-point function and its spectral representations at finite 
temperature in the real-time formalism. We then use these to 
explicitly calculate the 3-point spectral densities for $\phi^3_6$
theory and relate the result to the case of hot QCD.  

\end{abstract}


\section{Introduction}
\label{sec1}

Spectral functions are essential and useful in finite temperature 
field theory \cite{LvW87,K89} because a large number of transport 
coefficients are given directly by them \cite{HST84,J93,WHZ96}.  
Furthermore, studying the spectral functions may help us to understand 
the quasi-particle structure of field theories at finite temperature 
as well as to identify the microscopic processes underlying their 
dynamics. In this paper we derive expressions for the spectral 
densities of the 3-point Green functions in finite temperature 
field theories within the Closed Time Path (CTP) formalism 
\cite{keld,chou,henning} and evaluate the spectral densities of the 
3-point function for $\phi^3$ theory using resummed propagators in the 
``hard thermal loop'' (HTL) approximation \cite{pisarski,altherr,taylor}.  

In the imaginary-time formalism (ITF) \cite{K89}, one obtains the 
spectral densities from the discontinuity of the Green functions 
across the real energy axis after performing an analytic continuation 
of the imaginary external energy variables to the real axis 
\cite{evan1}.  Explicit expressions for the spectral densities for the 
three gluon ITF vertex in QCD in HTL approximation were derived in 
Ref.~\cite{taylor}.  

In real-time formulations of finite temperature field theory the 
number of degrees of freedom is doubled, leading to a $2\times 2$ 
matrix structure of the single particle propagators. The external 
energies remain real, and the complicated summation over the Masubara 
frequencies followed by analytic continuation is avoided. In 
Ref.~\cite{kob} Kobes and Semenoff derived Cutkosky rules for 
calculating the imaginary parts of thermal two-point functions using 
the formalism of Thermo-Field Dynamics (TFD). Spectral representations 
of the 3-point Green functions were derived in \cite{kob1} using 
the notation of ``circled'' vertices. Recently these cutting rules 
were reexamined in the CTP formalism in Ref.~\cite{BDN96,Gelis} and 
given a simple physical interpretation in Ref.~\cite{L96}.  

A useful technical simplification for perturbative calculations in 
real-time finite temperature field theory is provided by the 
decomposition and spectral representation of the 3-point vertex 
given in Ref.~\cite{CH96}. Missing in that paper is an explicit 
expression of the spectral densities in terms of the thermal 
components of the real-time 3-point vertex function. This hole is 
filled in by the present paper.

We then apply these expressions to the 3-point vertex for $\phi^3$ 
theory in 6 dimensions. We evaluate the corresponding spectral 
densities explicitly in the 1-loop approximation. It is well-known
\cite{K89,pisarski} that field theories with massless degrees of freedom 
develop at non-zero temperature infrared divergences which usually signal
dynamical mass generation and in many cases can be dealt with by resummation
of the ``hard thermal loops'' \cite{pisarski}. In $\phi^3$ theory the 
situation is even a little more complicated: the effective potential is
unbounded from below, and in the massless limit the theory doesn't even have 
a metastable ground state. By adding to the Lagrangean a non-zero, positive 
mass term the theory develops a metastable, local minimum at $\langle \phi
\rangle=0$ which, at zero temperature, is perturbatively stable in the 
limit of small 3-point coupling constant $g$ \cite{altherr}. At non-zero
temperature, however, the tadpole diagram contains a temperature dependent
finite, but negative contribution which shifts the position $\langle \phi 
\rangle$ of the metastable vacuum to negative values and reduces the 
effective boson mass \cite{altherr}. This effect must be taken into account
self-consistently via resummed (massive) propagators in order to avoid 
an expansion around the wrong vacuum.

We will use the CTP formalism \cite{keld,chou} throughout this paper 
in the form given in Refs.~\cite{henning,CH96}. In this representation of 
the real-time formalism the single-particle propagator in momentum space 
has the form 
 \begin{equation}
 \label{1}
   D(p) = \left(  \matrix {D_{11} & D_{12} \cr
                           D_{21} & D_{22} \cr} \right) \, ,
 \end{equation}
where
 \begin{mathletters}
 \label{2}
 \begin{eqnarray}
  i\, D_{11}(p) &=& \left(i\,D_{22}\right)^*
  = i{\cal P} \left(\frac {1}{p^2-m^2}\right) 
  + \left( n(p_0)+{1\over 2} \right) \rho(p)\, ,
 \label{2a} \\
  i\, D_{12}(p) &=&  n(p_0)\, \rho(p) \, ,
 \label{2b}\\
  i\, D_{21}(p) &=& 
  \bigl(1 + n(p_0) \bigr)\, \rho(p) \, .
 \label{2d}
 \end{eqnarray}
 \end{mathletters}
Here $n(p_0)$ is the thermal Bose-Einstein distribution
 \begin{equation}
 \label{3}
   n(p_0) = \frac{1}{ e^{\beta p_0}-1},
 \end{equation}
and $\rho(p)$ is the two-point spectral density which for free particles
is given by
 \begin{equation}
 \label{3a}
   \rho(p)= 2\pi\, {\rm sgn}(p_0) \,\delta(p^2-m^2)\, .
 \end{equation}

The paper is organized as follows. In Sec.~\ref{sec2} we review some 
useful general relations among the thermal components of the 3-point 
function and their spectral representations, both for the connected
and for the truncated vertices. In Sec.~\ref{sec3} we evaluate the 
spectral densities for the truncated 3-point vertex in $\phi^3$ 
in 1-loop approximation. In Sec.~\ref{sec4} we discuss and summarize
our results. Some technical details of the calculations and further
useful relations are given in the Appendix.   

\section{Spectral representation of the 3-point vertex}
\label{sec2}

In this Section we shortly review some useful relations among the
different thermal components of the 3-point functions and their
spectral representation. Equivalent (although not identical) relations 
have been reported in the literature \cite{chou,kob,kob1,evan2,kob2} 
in different notation. For simplicity of presentation we consider the 
3-point vertex function for $\phi^3$ theory, see Fig.~1. 
The three incoming external momenta are $k_1=p$, $k_2=q$, and $k_3=-p-
q$.  

\begin{eqnarray}
\parbox{14cm}
{{
\begin{center}
\parbox{10cm}
{
\epsfxsize=7cm
\epsfysize=6cm
\epsfbox{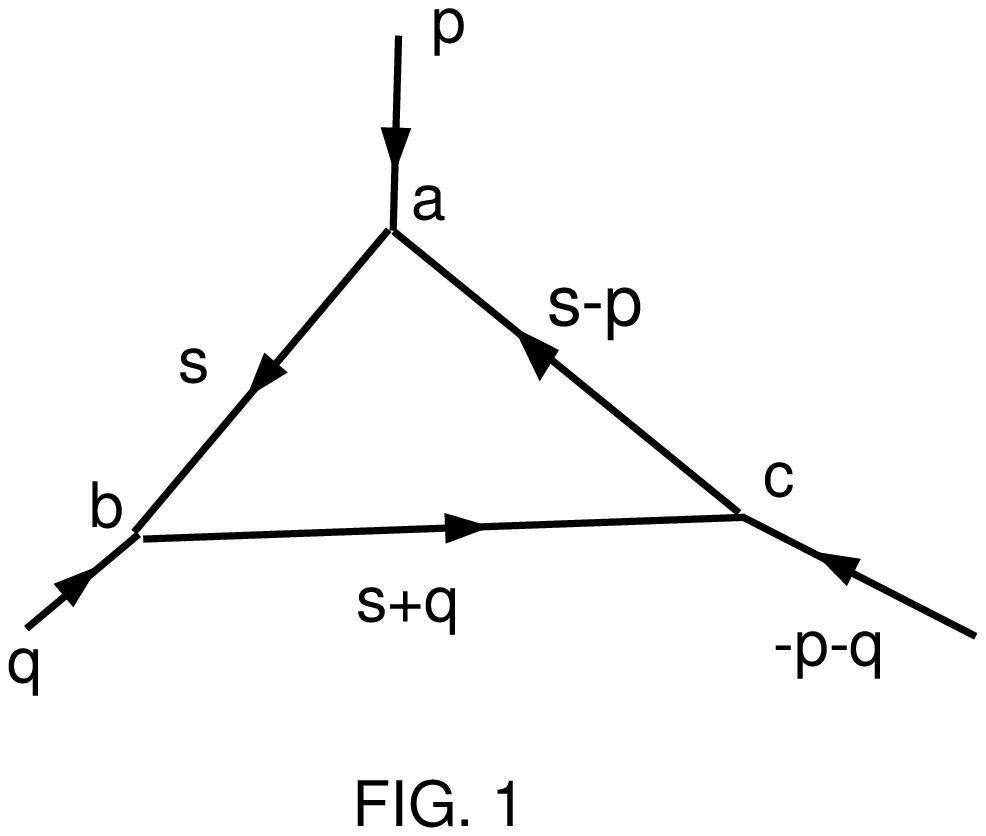}}\\
\parbox{14cm}{\small \center  Fig.~1: 3-point vertex in $\phi^3$ theory}
\end{center}
}}
\nonumber
\end{eqnarray}

\subsection{Relations among the thermal components of the real-time 
vertex}
\label{sec2a}

The thermal components of the connected 3-point vertex function
are defined by \cite{chou}
 \begin{mathletters}
 \label{4}
 \begin{eqnarray}
 \label{4a}
   \Gamma_{111} &=& \langle T(\phi_1\phi_2\phi_3) \rangle  \, ,
 \\
 \label{4b}  
   \Gamma_{112} &=& \langle \phi_3 T(\phi_1\phi_2) \rangle \, ,
 \\
 \label{4c}  
   \Gamma_{121} &=& \langle \phi_2 T(\phi_1\phi_3) \rangle \, ,
 \\
 \label{4d}  
   \Gamma_{211} &=& \langle \phi_1 T(\phi_2\phi_3) \rangle \, ,
 \\
 \label{4e}  
   \Gamma_{122} &=& \langle \tilde T(\phi_2\phi_3)\phi_1 \rangle \, ,
 \\
 \label{4f}
   \Gamma_{212} &=& \langle \tilde T(\phi_1\phi_3)\phi_2 \rangle \, ,
 \\
 \label{4g}
   \Gamma_{221} &=& \langle \tilde T(\phi_1\phi_2)\phi_3 \rangle \, ,
 \\
 \label{4h}
   \Gamma_{222} &=& \langle \tilde T(\phi_1\phi_2\phi_3) \rangle \, ,
 \end{eqnarray}
 \end{mathletters}
where $\phi_1 \equiv \phi(x_1) = \phi(\bbox{x}_1,t_1)$ etc., and 
$\Gamma_{abc} \equiv \Gamma_{abc}(x_1,x_2,x_3)$). Following 
Ref.~\cite{Umez} we defined the process of ``tilde conjugation'' 
by reversing the time order in coordinate space: time-ordered
products become products with anti-chronological ordering, 
and $\theta(t)$ becomes $\theta(-t)$.

Using the identity $\theta(t) + \theta(-t) = 1$ it is straightforward 
to show that 
 \begin{equation}
 \label{5}
   \sum_{a,b,c=1}^2 (-1)^{a+b+c-3}\Gamma_{abc} = 0 \, .
 \end{equation}
In momentum space tilde conjugation turns out to be equivalent to 
complex conjugation and, using the KMS condition, one finds \cite{chou,kob2}
 \begin{mathletters}
 \label{6}
 \begin{eqnarray}
 \label{6a}
   \tilde\Gamma_{111}(k_1,k_2,k_3) &=& \Gamma_{111}^*(k_1,k_2,k_3)
   = \Gamma_{222} (k_1,k_2,k_3) \, ,
 \\
 \label{6b}
   \tilde\Gamma_{121}(k_1,k_2,k_3) &=& \Gamma_{121}^*(k_1,k_2,k_3)
   = e^{\beta \omega_2}\,\Gamma_{212}(k_1,k_2,k_3) \, ,
 \\
 \label{6c}
   \tilde\Gamma_{211}(k_1,k_2,k_3) &=& \Gamma_{211}^*(k_1,k_2,k_3)
   = e^{\beta \omega_1}\,\Gamma_{122}(k_1,k_2,k_3) \, ,
 \\
 \label{6d}
   \tilde\Gamma_{112}(k_1,k_2,k_3) &=& \Gamma_{112}^*(k_1,k_2,k_3) 
   = e^{\beta\omega_3}\, \Gamma_{221}(k_1,k_2,k_3)
 \, ,
 \end{eqnarray}
 \end{mathletters}
where $k_i = (\omega_i,\bbox{k}_i)$ and $k_1+k_2+k_3=0$. These 
identities show that at most three of the eight thermal components of 
the real-time vertex function are independent.  

\subsection{Largest and smallest time equations}
\label{sec2b}

If $t_3$ is the largest time argument, one obtains from (\ref{4a}) and 
(\ref{4b}) the identities
 \begin{equation}
 \label{7}
   \theta_{32} \, \theta_{21} \, \Gamma_{111}
   = \langle \phi_3\phi_2\phi_1 \rangle =
   \theta_{32} \, \theta_{21} \, \Gamma_{112}
 \end{equation}
or
 \begin{equation}
 \label{8}
   \theta_{32} \, \theta_{21}\,(\Gamma_{111}-\Gamma_{112}) = 0 \, .
 \end{equation}
Here $\theta_{ij} \equiv \theta(t_i-t_j)$. Similarly one derives the 
more general relations
 \begin{mathletters}
 \label{9}
 \begin{eqnarray}
 \label{9a}
   \theta_{32}\,\theta_{21}\,(\Gamma_{ab1}-\Gamma_{ab2}) &=& 0 =
   \theta_{31}\,\theta_{12}\,(\Gamma_{ab1}-\Gamma_{ab2}) \, ,
 \\
 \label{9b}
   \theta_{21}\,\theta_{13}\,(\Gamma_{a1b}-\Gamma_{a2b}) &=& 0 =
   \theta_{23}\,\theta_{31}\,(\Gamma_{a1b}-\Gamma_{a2b}) \, ,
 \\
 \label{9c}
   \theta_{13}\,\theta_{32}\,(\Gamma_{1ab}-\Gamma_{2ab}) &=& 0 =
   \theta_{12}\,\theta_{23}\,(\Gamma_{1ab}-\Gamma_{2ab}) \, ,
 \end{eqnarray}
 \end{mathletters}
where $a$ and $b$ can be either 1 or 2. By tilde conjugation one
obtains from these equations the following relations:
 \begin{mathletters}
 \label{10}
 \begin{eqnarray}
 \label{10a}
   \theta_{12}\,\theta_{23}\,(\tilde\Gamma_{ab1}-\tilde\Gamma_{ab2}) &=& 0 =
   \theta_{21}\,\theta_{13}\,(\tilde\Gamma_{ab1}-\tilde\Gamma_{ab2}) \, ,
 \\
 \label{10b}
   \theta_{31}\,\theta_{12}\,(\tilde\Gamma_{a1b}-\tilde\Gamma_{a2b}) &=& 0 =
   \theta_{13}\,\theta_{32}\,(\tilde\Gamma_{a1b}-\tilde\Gamma_{a2b}) \, ,
 \\
 \label{10c}
   \theta_{23}\,\theta_{31}\,(\tilde\Gamma_{1ab}-\tilde\Gamma_{2ab}) &=& 0 =
   \theta_{32}\,\theta_{21}\,(\tilde\Gamma_{1ab}-\tilde\Gamma_{2ab}) \, ,
 \end{eqnarray}
 \end{mathletters}
Eqs.~(\ref{9}) and (\ref{10}) are the analogues of the ``largest time 
equations'' and ``smallest time equations'', respectively, of 
Ref.~\cite{kob}. They will be used extensively in the derivation of the 
spectral representations of Appendix~\ref{appa}. Their generalization 
to arbitrary $n$-point functions is straightforward.

\subsection{Physical vertex functions}
\label{sec2c}                                 

One can construct the ``retarded'', ``forward'', and ``even'' 
vertex functions from the eight components of the 3-point function 
as \cite{chou,CH96}.  
 \begin{mathletters}
 \label{11}
\begin{eqnarray}
 \label{11a}
  \Gamma_{R} &=& \Gamma_{111}-\Gamma_{112}-\Gamma_{211}+\Gamma_{212},
\\
 \label{11b}
  \Gamma_{Ri} &=& \Gamma_{111}-\Gamma_{112}-\Gamma_{121}+\Gamma_{122},
\\
 \label{11c}
  \Gamma_{Ro} &=& \Gamma_{111}-\Gamma_{121}-\Gamma_{211}+\Gamma_{221},
\\
 \label{11d}
  \Gamma_{F} &=& \Gamma_{111}-\Gamma_{121}+\Gamma_{212}-\Gamma_{222}, 
\\
 \label{11e}
  \Gamma_{Fi} &=& \Gamma_{111}+\Gamma_{122}-\Gamma_{211}-\Gamma_{222},
\\
 \label{11f}
  \Gamma_{Fo} &=& \Gamma_{111}-\Gamma_{112}+\Gamma_{221}-\Gamma_{222}, 
\\
 \label{11g}
  \Gamma_{E} &=& \Gamma_{111}+\Gamma_{122}+\Gamma_{212}+\Gamma_{221}, 
\end{eqnarray}
 \end{mathletters}
Inversion of these equations together with Eq.~(\ref{5}) yields
expressions for the thermal components $\Gamma_{abc}$ in terms 
the above ``physical'' vertex functions; they are given in compact 
form in Ref.~\cite{CH96}. 

Using Eqs.~(\ref{5}) and (\ref{6}) one can eliminate the ``forward''
and ``even'' vertex functions in terms of the three retarded
vertices. Thus all components of $\Gamma_{abc}$ can be expressed
through $\Gamma_{R}$, $\Gamma_{Ri}$, and $\Gamma_{Ro}$
\cite{evan2,CH96}.   

\subsection{Spectral integral representations}
\label{sec2d}                                 
          
In Ref.~\cite{CH96} the following integral representations for the
retarded vertex functions in momentum space were derived (in 
slightly different notation):
 \begin{mathletters}
 \label{12}
 \begin{eqnarray}
 \label{12a}
   \Gamma_{R}(\omega_1,\omega_2,\omega_3) &=&
   \frac{-i}{2\pi^2} \int_{-\infty}^{\infty}
   \frac{d\Omega_1 d\Omega_2}{\omega_2-\Omega_2+i\epsilon}
   \left(
   \frac{\rho_1}{\omega_1-\Omega_1-i\epsilon}
  +\frac{\rho_1-\rho_2}{\omega_3-\Omega_3-i\epsilon} \right) \, ,
 \\
 \label{12b}
   \Gamma_{Ri}(\omega_1,\omega_2,\omega_3) &=&
   \frac{-i}{2\pi^2} \int_{-\infty}^{\infty}
   \frac{d\Omega_1 d\Omega_2}{\omega_1-\Omega_1+i\epsilon}
   \left(
   \frac{\rho_2}{\omega_2-\Omega_2-i\epsilon}
  -\frac{\rho_1-\rho_2}{\omega_3-\Omega_3-i\epsilon} 
   \right)\, ,
 \\
 \label{12c}
   \Gamma_{Ro}(\omega_1,\omega_2,\omega_3) &=&
   \frac{-i}{2\pi^2}\int_{-\infty}^{\infty}
   \frac{d\Omega_1 d\Omega_2}{\omega_3-\Omega_3+i\epsilon}
   \left(   
   \frac{\rho_1}{\omega_1-\Omega_1-i\epsilon}
  +\frac{\rho_2}{\omega_2-\Omega_2-i\epsilon} \right)\, .
 \end{eqnarray}
 \end{mathletters}
The spatial momenta $\bbox{k}_1$, $\bbox{k}_2$, $\bbox{k}_3{=}{-}
(\bbox{k}_1{+}\bbox{k}_2)$ are the same on both sides of these equations 
and have therefore been suppressed. The frequency arguments 
of the spectral functions under the integrals are $\rho_i \equiv 
\rho_i(\Omega_1, \Omega_2, \Omega_3)$, with $\Omega_1 + \Omega_2 + 
\Omega_3 = 0$. In Appendix~\ref{appa} we give a short derivation of 
these integral representations from which it follows that in momentum
space 
 \begin{mathletters}
 \label{14}
 \begin{eqnarray}
 \label{14a}
   \rho_1 &=& {\rm Im\,} (\Gamma_{122} + \Gamma_{211}) \, ,
 \\
 \label{14b}
   \rho_2 &=& {\rm Im\,} (\Gamma_{121} + \Gamma_{212}) \, .
 \end{eqnarray}
 \end{mathletters}

The spectral integral representations (\ref{12}) differ from those
given in Eqs.~(31) of Ref.~\cite{CH96} because they use different 
spectral densities. The spectral functions $\rho_1$, $\rho_2$ used here 
are not simply related to $\rho_A$, $\rho_B$ of Ref.~\cite{CH96}: while
it follows from Eqs.~(\ref{14}) that $\rho_1$ and $\rho_2$ are real 
in momentum space, $\rho_A$ and $\rho_B$ are instead real in coordinate 
space and satisfy a more complicated relation (Eq.~(28) of 
Ref.~\cite{CH96}) in momentum space. Still, both sets of spectral 
integral representations are correct; the one given here appears to 
simplify things in practice, however (see below). 

Similar spectral representations can be derived for the truncated
(1-particle irreducible, 1PI) vertex functions. The technical steps 
are given in Appendix~\ref{appa2}, together with the corresponding
generalizations to 1PI vertex functions for the relations derived in
the preceding subsections. Here we only state the result: 
 \begin{mathletters} 
 \label{c21}
 \begin{eqnarray}
 \label{c21a}
   G_{R}(\omega_1,\omega_2,\omega_3) &=&
   \frac{-i}{2\pi^2} \int_{-\infty}^{\infty}
   \frac{d\Omega_1 d\Omega_2}{\omega_2-\Omega_2+i\epsilon}
   \left(
   \frac{\rho'_1}
        {\omega_1-\Omega_1-i\epsilon}
  +\frac{\rho'_1-\rho'_2}
        {\omega_3-\Omega_3-i\epsilon} \right) \, ,
 \\
 \label{c21b}
   G_{Ri}(\omega_1,\omega_2,\omega_3) &=&
   \frac{-i}{2\pi^2} \int_{-\infty}^{\infty}
   \frac{d\Omega_1 d\Omega_2}{\omega_1-\Omega_1+i\epsilon}
   \left(
   \frac{\rho'_2}
        {\omega_2-\Omega_2-i\epsilon} 
  -\frac{\rho'_1-\rho'_2}
        {\omega_3-\Omega_3-i\epsilon} \right)\, ,
 \\
 \label{c21c}
   G_{Ro}(\omega_1,\omega_2,\omega_3) &=&
   \frac{-i}{2\pi^2}\int_{-\infty}^{\infty}
   \frac{d\Omega_1 d\Omega_2}{\omega_3-\Omega_3+i\epsilon}
   \left(
   \frac{\rho'_1}
        {\omega_1-\Omega_1-i\epsilon} 
  +\frac{\rho'_2}
        {\omega_2-\Omega_2-i\epsilon} \right)\, , 
 \end{eqnarray} 
 \end{mathletters} 
with 
 \begin{mathletters}
 \label{c19}
 \begin{eqnarray}
 \label{c19a}
   \rho'_1 &=& {\rm Im\, } (G_{122} - G_{211}) \, ,
 \\
 \label{c19b}
   \rho'_2 &=& {\rm Im\, } (G_{212} - G_{212}) \, .
 \end{eqnarray}
 \end{mathletters}

These results should be compared with the expressions derived by Kobes
in Ref.~\cite{kob1} which have a similar structure but use three
somewhat differently defined spectral densities.

\section{1-loop spectral densities for the vertex in $\phi^3_6$ theory} 
\label{sec3}

In this Section we calculate the 1-loop contribution to the spectral
functions $\rho'_1$, $\rho'_2$ for the 1PI 3-point vertex in
$\phi^3$ theory. As interaction term in the Lagrangean we use ${g\over
  6} \phi^3$. From Eqs.~(\ref{c19}) and (\ref{c3}) we have  
 \begin{equation}
 \label{15}
   \rho_1'  = {\rm Im\,} (G_{122} + e^{\beta p_0}G_{122}^*)
   = - {1\over n(p_0)} {\rm Im\, } G_{122} \, .
 \end{equation}
For $\rho'_1$ we thus must evaluate only the single Feynman diagram 
in Fig.~1 for $a=1$, $b=c=2$. Using standard real-time Feynman rules
\cite{chou,henning} one gets (in $n$ space-time dimensions) 
 \begin{equation}
 \label{16}
   G_{122}(p,q,-p-q)=(-ig)(ig)^2\int \frac {d^n s}{(2\pi)^n} \,
   [iD_{12}(s)] [iD_{22}(s+q)] [iD_{21}(s-p)] \, .
 \end{equation}
Inserting the thermal free propagators (\ref{2}), extracting the  
imaginary part, and performing the integration over $s^0$ with the help
of the function $\delta(s^2-m^2) = [\delta(s_0-E_s) + \delta(s_0+E_s)]/2E_s$,
where $E_s = \sqrt{m^2+{\bbox{s}^2}}$, one finds 
 \begin{mathletters}
 \label{17}
 \begin{eqnarray}
 \label{17a}
   \rho_1(p,q,-p-q) &=& 
   - {g^3 \over n(p_0)} \Bigl( A(p,q) + B(p,q) \Bigr)\, ,
 \\
 \label{17b}
   A(p,q) &=& \int {d^{n-1}s\over (2\pi)^{n-3}} \, {1\over 2 E_s}
          {\rm sgn}(E_s+q_0)\, {\rm sgn}(E_s-p_0 )\, 
 \nonumber\\
   && \ \ \  \times \, 
      \delta\bigl( (E_s+q_0)^2 - E_{s+q}^2 \bigr) \,
      \delta\bigl( (E_s-p_0)^2 - E_{s-p}^2 \bigr) 
 \nonumber\\
   && \ \ \ \times \,
      n(E_s) \, \Bigl( {\textstyle{1\over 2}} + n(E_s+q_0) \Bigr) 
      \, \bigl( 1 + n(E_s-p_0) \bigr)\, ,
 \\
 \label{17c}
   B(p,q) &=& \int {d^{n-1}s\over (2\pi)^{n-3}} \, {1\over 2 E_s}
          {\rm sgn}(E_s-q_0)\, {\rm sgn}(E_s+p_0 )\, 
 \nonumber\\
   && \ \ \ \times 
      \delta\bigl( (E_s-q_0)^2 - E_{s+q}^2 \bigr) \,
      \delta\bigl( (E_s+p_0)^2 - E_{s-p}^2 \bigr) 
 \nonumber\\
   && \ \ \ \times 
      \bigl( 1+n(E_s) \bigr) \, 
      \Bigl( {\textstyle{1\over 2}} + n(E_s-q_0) \Bigr) \, 
      n(E_s+p_0)\, .
 \end{eqnarray}
 \end{mathletters}
Here $E_{s+q} = \sqrt{m^2 + (\bbox{s}+\bbox{q})^2}$, $E_{s-p} = 
\sqrt{m^2 + (\bbox{s}-\bbox{p})^2}$, and in (\ref{17c}) we used the 
identity $n(-x) = -\bigl(1+n(x)\bigr)$. 

The integrands in Eqs.~(\ref{17}b,c) contain up to three powers of the
thermal Bose distribution functions. Superficial power counting thus
suggests severe infrared singularities in the massless limit. On the
other hand, 1-loop integrals in the imaginary time formalism are
always linear in the thermal distribution functions which arise from
conversion of a single sum over discrete loop frequencies into a complex
contour integral \cite{pisarski}. This suggests that the infrared
problems resulting from higher powers of the distribution functions in
the real time formalism are spurious. In fact, it was already noted in
Refs.~\cite{bair,ashida} that the cubic terms cancel from the retarded
3-point functions. Using the identity
 \begin{equation}
 \label{reldis}
   n(a)\, n(b) = n(a+b)\, \bigl(1+n(a)+n(b)\bigr)
 \end{equation}
one can show that also the quadratic terms disappear, and that $A$ and
$B$ reduce to
 \begin{mathletters}
 \label{1+}
 \begin{eqnarray}
 \label{1a+}
   A(p,q) &=& n(p_0) \int {d^{n-1}s\over (2\pi)^{n-3}} \, {1\over 2 E_s}
          {\rm sgn}(E_s+q_0)\, {\rm sgn}(E_s-p_0 )\, 
 \nonumber\\
   && \qquad \times \, 
      \delta\bigl( (E_s+q_0)^2 - E_{s+q}^2 \bigr) \,
      \delta\bigl( (E_s-p_0)^2 - E_{s-p}^2 \bigr) 
 \nonumber\\
   && \qquad \times \,
      \Bigl[ {\textstyle{1\over 2}}\bigl(n(E_s-p_0)-n(E_s)\bigr)
              +n(p_0+q_0)\bigl(n(E_s-p_0)-n(E_s+q_0)\bigr)
 \nonumber\\
   && \qquad \quad 
             +n(q_0)\bigl(n(E_s+q_0)-n(E_s)\bigr) \Bigr] \, ,
 \\
 \label{1b+}
   B(p,q) &=& n(p_0) \int {d^{n-1}s\over (2\pi)^{n-3}} \, {1\over 2 E_s}
          {\rm sgn}(E_s-q_0)\, {\rm sgn}(E_s+p_0 )\, 
 \nonumber\\
   && \qquad \times \,
      \delta\bigl( (E_s-q_0)^2 - E_{s+q}^2 \bigr) \,
      \delta\bigl( (E_s+p_0)^2 - E_{s-p}^2 \bigr) 
 \nonumber\\
   && \qquad \times \,
      \Bigl[ {\textstyle{1\over 2}}\bigl(n(E_s+p_0)-n(E_s)\bigr)
            +n(p_0+q_0)\bigl(n(E_s+p_0)-n(E_s-q_0)\bigr)
 \nonumber\\
   && \qquad \quad 
            +n(q_0)\bigl(n(E_s-q_0)-n(E_s)\bigr) \Bigr]
 \end{eqnarray}
 \end{mathletters}
Note that the factor $n(p_0)$ in front of the integrals cancels the
distribution function in the denominator of Eqs.~(\ref{15}),
(\ref{17a}). The remaining integrands are linear in the thermal
distribution functions $n(E_s\pm \dots)$ and are infrared finite even
in the massless limit \cite{Aur}.

To simplify the notation it is convenient to introduce the 4-vectors
$V_\pm = \left(1, \pm {\bbox{s}\over E_s} \right)$. With their help we
can rewrite the arguments of the $\delta$-functions in (\ref{17}) as 
 \begin{mathletters}
 \label{18}
 \begin{eqnarray}
 \label{18a}
    \left. (s+q)^2 \right\vert_{s^0=\pm E_s} 
    &=& m^2+q^2 \pm 2E_s\, q\cdot V_\pm \, ,
 \\     
 \label{18b}
    \left. (s-p)^2 \right\vert_{s^0=\pm E_s} 
    &=& m^2+p^2 \mp 2E_s\, p\cdot V_\pm \, .
 \end{eqnarray}
 \end{mathletters}

We will consider the theory in the weak coupling limit, $g \ll 1$.
As shown in \cite{altherr}, for $n=6$ the theory becomes
perturbatively unstable\footnote{Stability problems for perturbation
theory in massless $\phi^3$ theory were recently also discussed in the
context of relativistic transport thery in Ref.~\cite{mrow}.} for
temperatures above 
 \begin{equation}
 \label{19}
   T_{\rm cr} = \left( {180\over \pi} \right)^{1/4} {m\over \sqrt{g}}\, ,
 \end{equation}
so for given $T$ we must use massive propagators with a mass that
satisfies $m > \sqrt{g}\, T$. We will consider the case $\sqrt{g}\, T
\lesssim m \ll T$ and calculate the spectral density for the vertex
for soft external momenta, $q,\, p \sim m \sim \sqrt{g}\, T \ll
T$. 

For gauge theories it is known that in the limit of soft external
momenta there is a ``hard thermal loop'' (HTL) contribution to the
3-gluon vertex which is of the same order in the coupling constant as
the tree-level result and must therefore be resummed in a complete
leading order calculation \cite{pisarski}. To see whether such a
resummation is also required in scalar $\phi^3$ theory we first
evaluate the functions $A,B$ in the hard thermal loop approximation,
by assuming that the loop integral in (\ref{18}) is dominated by
``hard'' momenta $\bar s = \vert \bbox{s} \vert \sim T$
\cite{pisarski}. For such momenta we can neglect the rest mass $m$,
$E_s \approx \bar s$, and the arguments of the $\delta$-functions can
be approximated according to    
 \begin{mathletters}
 \label{20}
 \begin{eqnarray}
 \label{20a}
    \left. (s+q)^2 \right\vert_{s^0=\pm E_s} &\approx&
    \pm 2E_s\, q\cdot V_\pm \, ,
 \\     
 \label{20b}
    \left. (s-p)^2 \right\vert_{s^0=\pm E_s} 
    &\approx& \mp 2E_s\, p\cdot V_\pm \, .
 \end{eqnarray}
 \end{mathletters}
Setting further 
 \begin{mathletters}
 \label{21}
  \begin{eqnarray}
 \label{21a}
  && {\rm sgn}(E_s \pm p^0) \approx 1 \approx {\rm sgn}(E_s \pm q^0) \, ,
 \\
 \label{21b}
  && q\cdot V_\pm \approx q^0 \mp \vert \bbox{q}\vert \cos\theta'\, ,
 \\     
 \label{21c}
  && p\cdot V_\pm \approx p^0 \mp \vert \bbox{p}\vert \cos\theta\, ,
 \end{eqnarray}
 \end{mathletters}
where $\theta$, $\theta'$ are the angles between $\bbox{s}$ and
$\bbox{p}$, $\bbox{q}$, respectively, the angular and radial
integrations in (\ref{1+}) decouple \cite{pisarski}. We thus obtain 
 \begin{mathletters}
 \label{22}
 \begin{eqnarray}
 \FL
 \label{22a}
   &&A(p,q) + B(p,q)\Big\vert_{\rm HTL} 
   = n(p_0)\, a(p_0,q_0) \, \omega(p,q)\, ,
 \\
 \label{22b}
   &&a(p_0,q_0) = {1 \over 2^n \pi^{n-3}} \int_0^\infty 
         d\bar s \, {\bar s}^{n-5} \Bigl[ {\textstyle{1\over 2}}
         \Bigl( n(\bar s-p_0) + (n(\bar s+p_0) - 2 n(\bar s)\Bigr)
 \nonumber\\
   && \qquad \qquad  
        + n(q_0) \Bigl( n(\bar s+q_0) + n(\bar s-q_0) - 2 n(\bar s)\Bigr)
 \\
   && \quad \quad 
        + n(p_0+q_0) \Bigl( n(\bar s+p_0) + n(\bar s-p_0)
                          - n(\bar s+q_0) - n(\bar s-q_0)\Bigr) \Bigr] \, ,
 \nonumber\\
 \label{22c}
   &&\omega(p,q) = 
   \int d\Omega_{n-1} \, \delta(q \cdot V_+) \, \delta(p \cdot V_+) =
   \int d\Omega_{n-1} \, \delta(q \cdot V_-) \, \delta(p \cdot V_-) \, .
 \end{eqnarray}
 \end{mathletters}
The angular integral (\ref{22c}) is identical with the one found by
Taylor \cite{taylor} for the spectral density of the 3-gluon vertex
in hot QCD. For $n=6$ space-time dimensions (for which the theory is
renormalizable) the radial integral (\ref{22b}) is easily evaluated
with the help of
 \begin{eqnarray}
 \label{B1}
   I(a) &=&  \int_0^\infty d\bar s \, \bar s 
   \left({1\over e^{\bar s+a}-1}+{1\over e^{\bar s-a}-1}\right)
 \nonumber\\
     &=& \int_0^a (x-a)\, dx 
         + 2 \int_0^\infty {x\, dx\over e^x-1}
     = {\pi^2\over 3} - {a^2\over 2} \, .
 \end{eqnarray}
We find
 \begin{equation}
 \label{B2}
   a(p_0,q_0) = - {1\over 2^7 \pi^3} \Bigl[ 
   p_0^2 \Bigl( {\textstyle{1\over 2}} + n(p_0+q_0) \Bigr) +
   q_0^2 \Bigl( n(q_0)-n(q_0+p_0) \Bigr)\Bigr]\, .
 \end{equation}
For $p_0, q_0 \ll T$ this goes to
 \begin{equation} 
 \label{B3}
   a(p_0,q_0) \approx - {1\over 2^7 \pi^3} p_0 T\, .
 \end{equation}
For $p_0 \sim \sqrt{g}\, T \ll T$ this is much smaller than the
leading $T^2$-behaviour expected on dimensional grounds; this implies
that the assumption that the loop integral is dominated by hard momenta
$\bar s \sim T$ was wrong, and that in scalar $\phi^3$ theory 
{\em there is no leading HTL contribution to the 3-point vertex}, in 
contrast to the case of gauge theories. Braaten-Pisarski resummation
for $\phi^3$ theory can thus be performed with bare 3-point vertices.
A similar result was obtained in Ref.~\cite{schulz,BP92} for scalar QED. 
The existence of leading HTL contributions to vertices in QCD and
fermionic QED can be traced back to the existence of Ward identities
which connect vertex corrections with self energy corrections
\cite{pisarski,BP92,FT92}. 

Before proceeding to a more accurate evaluation of the spectral
density for vanishing spatial external momenta, let us shortly comment
on the other spectral density which is obtained from
 \begin{equation}
 \label{25}
   \rho_2' = -{1\over n(q_0)} {\rm Im}G_{212} \, .
 \end{equation}
By inspection of the corresponding labelling of the diagram in Fig.~1
one observes that $G_{212}(p,q,-p-q)$ is obtained from
$G_{122}(p,q,-p-q)$ by exchanging the legs with the external momenta
$p$ and $q$ and routing the internal momentum $s$ in the opposite
direction. This yields the identity\footnote{Note that such an
identity is not expected to hold for the 3-point vertex in Yukawa
theory or in other theories where different types of fields are
attached to the vertex.} 
 \begin{equation}
 \label{25a}
  \rho'_2(p,q) = \rho'_1(q,p)\, .
 \end{equation}
For the HTL contributions to the corresponding loop integrals we thus
obtain 
 \begin{mathletters}
 \label{26}
 \begin{eqnarray}
 \label{26a}
    \rho_1^{\rm HTL}(p,q) &=& p_0\, \rho_{\rm HTL}(p,q) \, ,
 \\
 \label{26b}
    \rho_2^{\rm HTL}(p,q) &=& q_0 \rho_{\rm HTL}(p,q) \, ,
 \\
    \rho_{\rm HTL}(p,q) &\approx& {g^3 T \over 2^7\pi^3}  
    \int d\Omega_5 \, \delta(p{\cdot}V_+)\, \delta(q{\cdot}V_+)\, .
 \end{eqnarray}
 \end{mathletters}
Up to a trivial external momentum factor the two spectral densities
are thus equal to each other in the HTL limit. This agrees with the
observation by Taylor \cite{taylor} that in QCD in HTL approximation
the two independent spectral densities for the 3-gluon vertex 
degenerate. 

The result (\ref{B3}) shows that the loop integral is {\em not}
dominated by hard momenta of order $T$, contrary to the assumption
under which the integral was evaluated. This means that the HTL result
for the spectral density is not reliable as an order of magnitude
estimate, not even for power counting in the coupling constant $g$. In
general a better estimate is difficult to obtain because for small
loop momenta the radial and angular integrations cannot be
decoupled. Things simplify, however, for vanishing external spatial 
momenta, $\bbox{q} = \bbox{p} =0$. In this limit we find
 \begin{mathletters}
 \label{2+}
 \begin{eqnarray}
 \label{2a+}
   A(p_0,q_0) &=& n(p_0) \int {d^{n-1}s\over (2\pi)^{n-3}} \, {1\over 2 E_s}\,
          {\rm sgn}(E_s+q_0)\, {\rm sgn}(E_s-p_0 )\, 
  {\delta (2E_s+q_0) \, \delta (2E_s-p_0)\over |p_0||q_0|} \,   
 \nonumber\\
   && \quad \quad \times \,
      \Bigl[ {\textstyle{1\over 2}} \bigl( n(E_s-p_0)-n(E_s)\bigr)
            - n(p_0+q_0)\bigl(n(E_s+q_0)-n(E_s-p_0)\bigr)
 \nonumber\\
   && \quad \qquad 
            + n(q_0)\bigl(n(E_s+q_0)-n(E_s)\bigr) \Bigr] \, ,
 \\
 \label{2+c}
   B(p_0,q_0) &=& n(p_0) \int {d^{n-1}s\over (2\pi)^{n-3}} \, {1\over 2 E_s}\,
          {\rm sgn}(E_s-q_0)\, {\rm sgn}(E_s+p_0 )\, 
  {\delta (2E_s-q_0) \, \delta (2E_s+p_0)\over |p_0||q_0|} \,   
 \nonumber\\
   && \quad \quad \times \,
      \Bigl[ {\textstyle{1\over 2}} \bigl( n(E_s+p_0)-n(E_s)\bigr)
            - n(p_0+q_0)\bigl(n(E_s-q_0)-n(E_s+p_0)\bigr)
 \nonumber\\
   && \quad \qquad 
            + n(q_0)\bigl(n(E_s-q_0)-n(E_s)\bigr) \Bigr] \, ,
 \end{eqnarray}
 \end{mathletters}
Due to the $\delta$-functions $p_0$ and $q_0$ must have the same
magnitude and opposite sign, and $A$ contributes only for $p_0 = -q_0
> 2m$ while $B$ contributes for $p_0 = -q_0 < -2m$. The angular
integrations are now trivial, and the radial integration is easily
performed using the $\delta$-functions. The final result, to leading
order in the small ratios $p_0/T,\, q_0/T$, is
 \begin{equation}
 \label{31}
  \rho'_1(p_0,q_0;\bbox{p}=\bbox{q} = 0) 
  \approx - {g^3\over 12\pi} \, {(p_0^2-4m^2)^{3/2} \over p_0^2 q_0^2}
    T^2\, \Bigl[ \theta(p_0-2m) - \theta(-p_0-2m) \Bigr] \, 
    \delta(p_0+q_0) \, .
 \end{equation}
Using (\ref{25a}) one obtains for the other spectral density
 \begin{equation}
 \label{32}
  \rho'_2(p_0,q_0;\bbox{p}=\bbox{q} = 0) =
  \rho'_1(q_0,p_0;\bbox{q}=\bbox{p} = 0) =
  - \rho'_1(p_0,q_0;\bbox{p}=\bbox{q} = 0)\, . 
 \end{equation}
For $p_0\sim q_0 \sim \sqrt{g}\, T$ power counting shows that
these spectral densities are of order $g^2$. Inserting them into
the spectral representations (\ref{c21}) and evaluating the latter via
residue calculus it is easy to see \cite{leupold} that the retarded
1-loop 1PI vertex functions at zero external spatial momenta are of
the same order, i.e. one order of $g$ down relative to the tree-level
vertex. This reconfirms the above conclusion that in scalar $\phi^3_6$
theory no vertex resummation is necessary. Note that this conclusion
depends crucially on the fact that the theory requires a (resummed)
mass of order $m\sim \sqrt{g}\, T$ for perturbative stability. If we
could take the limit $m\to 0$ and consider smaller external momenta of
order $p,q \sim gT$, naive power counting would give a 1-loop result
of order $g$, i.e. of the same order as the tree-level
vertex. However, in this limit the theory is not well-defined 
\cite{altherr,mrow}. 

\section{Conclusions}
\label{sec4}

In the CTP approach we have derived a set of useful relations among 
the eight thermal components of the 3-point vertex function many of
which we have not previously seen in the literature in this form. 
They simplify formal manipulations in the real-time formulation of 
finite temperature field theory. With their help we have found an
alternative derivation of spectral representations, in terms of two
independent spectral densities, for the various thermal components of 
the real-time 3-point vertex at finite temperature; they appear simpler 
than those given in the literature before.

We then proceeded to an explicit evaluation of these two spectral 
densities for the 3-point vertex in hot $\phi^3$ theory in 5+1
dimensions, in the 1-loop approximation for soft external momenta
$p,q \sim \sqrt{g}\, T$. This scale is set by the value of the (resummed)
scalar mass $m$ which is required to render the vacuum in $\phi^3_6$
theory perturbatively stable. We found that, contrary to the case of the
3-gluon vertex in QCD, the loop integral for the spectral density for
the scalar 3-point vertex is not dominated by hard momenta of order
$T$, and the popular HTL approximation which decouples the radial and
angular integrals produces an unreliable result. On the other hand,
this means that even for soft external momenta the 1-loop vertex is of
lower order than the tree-level contribution, and no vertex
resummation is necessary in the Braaten-Pisarski high-temperature
resummation scheme. An explicit evaluation of the 1-loop spectral
densities for the 3-point vertex at vanishing external spatial momenta
yields a result which is of order $g^2$, one power of $g$ (but not two
powers of $g$ as in naive perturbation theory) below the tree-level
vertex.

We also showed that the two independent spectral densities for the
3-point vertex in $\phi^3$ theory are very closely related by the
simple symmetry relation (\ref{25a}). At vanishing external spatial
momenta they become, up to a sign, equal to each other. This should be
compared with the finding of Taylor \cite{taylor} that in QCD in HTL
approximation the two spectral densities for the 3-gluon vertex become
identical. 

\acknowledgments

We are grateful to R.D. Pisarski for helpful comments on $\phi^3_6$ 
theory at finite temperature and to S. Leupold for insightful remarks
on the need for vertex resummation and for pointing out an
error in the original manuscript. This  work was  supported by the 
Deutsche Forschungsgemeinschaft (DFG), the Bundesministerium f\"ur
Bildung und Forschung (BMBF), the National Natural Science Foundation of
China (NSFC), and the Gesellschaft f\"ur Schwerionenforschung (GSI).


\appendix
\section{Derivation of spectral representations}
\label{appa}
\subsection{Connected three-point functions}
\label{appa1}

We start from the explicit expressions for the retarded connected 3-point 
vertex functions in $\phi^3$ theory:
 \begin{mathletters}
 \label{A1}
 \begin{eqnarray}
   \Gamma_R &=& 
     \theta_{23} \theta_{31} \langle[[\phi_2,\phi_3],\phi_1]\rangle
    +\theta_{21} \theta_{13} \langle[[\phi_2,\phi_1],\phi_3]\rangle \, ,
 \label{A1a}\\
   \Gamma_{Ri} &=& 
     \theta_{12} \theta_{23} \langle[[\phi_1,\phi_2],\phi_3]\rangle
    +\theta_{13} \theta_{32} \langle[[\phi_1,\phi_3],\phi_2]\rangle \, ,
 \label{A1b}\\
   \Gamma_{Ro} &=& 
     \theta_{32} \theta_{21} \langle[[\phi_3,\phi_2],\phi_1]\rangle
    +\theta_{31} \theta_{12} \langle[[\phi_3,\phi_1],\phi_2]\rangle \, .
 \label{A1c}
 \end{eqnarray}
 \end{mathletters}
{\bf (i)} We begin with $\Gamma_{Ro}$. Inserting the identities
 \begin{equation}
 \label{A2}
   \theta_{31} \Gamma_{Ro} = \Gamma_{Ro} = \theta_{32} \Gamma_{Ro} 
 \end{equation}
into 
 \begin{equation}
 \label{A3}
   \Gamma_{Ro} = (\theta_{21} + \theta_{12}) \Gamma_{Ro} 
 \end{equation}
one obtains
 \begin{equation}
 \label{A4}
   \Gamma_{Ro} = \theta_{31} \theta_{12} \Gamma_{Ro} 
               + \theta_{32} \theta_{21} \Gamma_{Ro}\, . 
 \end{equation}
Subtraction of the identities 
 \begin{equation}
 \label{A5}
   \theta_{31} \theta_{12} \Gamma_{R} = 0 = 
   \theta_{32} \theta_{21} \Gamma_{Ri} \, ,
 \end{equation}
which result from conflicting $\theta$-functions, then gives
 \begin{eqnarray}
 \label{A6}
   \Gamma_{Ro} &=& \theta_{32} \theta_{21} (\Gamma_{Ro} - \Gamma_{Ri}) 
                 + \theta_{31} \theta_{12} (\Gamma_{Ro} - \Gamma_{R})
 \nonumber\\
   &=& \theta_{32} \theta_{21} (\Gamma_{221} + \Gamma_{112} - 
                                \Gamma_{211} - \Gamma_{122})
     + \theta_{31} \theta_{12} (\Gamma_{221} + \Gamma_{112} - 
                                \Gamma_{121} - \Gamma_{212}) \, . 
 \end{eqnarray}
With the help of Eqs.~(\ref{9a}) and (\ref{6a}) this is transformed into
 \begin{eqnarray}
 \label{A7}
   \Gamma_{Ro} 
   &=& \theta_{32} \theta_{21} (\Gamma_{222} + \Gamma_{111} - 
                                \Gamma_{211} - \Gamma_{122})
     + \theta_{31} \theta_{12} (\Gamma_{222} + \Gamma_{111} - 
                                \Gamma_{121} - \Gamma_{212}) 
 \nonumber\\
   &=& \theta_{32} \theta_{21} (\tilde\Gamma_{111} + \tilde\Gamma_{222} - 
                                \Gamma_{211} - \Gamma_{122})
     + \theta_{31} \theta_{12} (\tilde\Gamma_{111} + \tilde\Gamma_{222} - 
                                \Gamma_{121} - \Gamma_{212}) 
 \, . 
 \end{eqnarray}
Using also Eqs.~(\ref{10}b,c) one finally gets
 \begin{equation}
 \label{A8}
   \Gamma_{Ro} = \theta_{32} \theta_{21} 
                 \bigl( \tilde\Gamma_{122} + \tilde\Gamma_{211} 
                     - (\Gamma_{122} + \Gamma_{211}) \bigr)
               + \theta_{31} \theta_{12} 
                 \bigl(\tilde\Gamma_{121} + \tilde\Gamma_{212} 
                     - (\Gamma_{121} + \Gamma_{212}) \bigr) \, .
 \end{equation}
{\bf (ii)} For $\Gamma_R$ one proceeds similarly. One writes
 \begin{equation}
 \label{A9}
   \Gamma_R = (\theta_{31}+\theta_{13}) \Gamma_R =
   \theta_{31} (\theta_{23} \Gamma_R) + \theta_{31} (\theta_{21} \Gamma_R)
 \end{equation} 
and subtracts the identities
 \begin{equation}
 \label{A10}
   \theta_{21} \theta_{13} \Gamma_{Ro} = 0 = 
   \theta_{23} \theta_{31} \Gamma_{Ri} \, .
 \end{equation}
This yields
 \begin{equation}
 \label{A11}
   \Gamma_{R} = \theta_{21} \theta_{13} \bigl(
   \Gamma_{121} + \Gamma_{212} - (\Gamma_{112} + \Gamma_{221}) \bigr)
              + \theta_{23} \theta_{31} \bigl(
   \Gamma_{212} + \Gamma_{121} - (\Gamma_{211} + \Gamma_{122}) \bigr)\, . 
 \end{equation}
Using Eqs.~(\ref{9b}) and (\ref{10}a,c) this is then transformed into
 \begin{equation}
 \label{A12}
   \Gamma_{R} = \theta_{21} \theta_{13} 
                 \bigl( \tilde\Gamma_{112} + \tilde\Gamma_{221} 
                     - (\Gamma_{112} + \Gamma_{221}) \bigr)
               + \theta_{23} \theta_{31} 
                 \bigl(\tilde\Gamma_{211} + \tilde\Gamma_{122} 
                     - (\Gamma_{211} + \Gamma_{122}) \bigr) \, .
 \end{equation}
{\bf (iii)} Finally, $\Gamma_{Ri}$ is reexpressed by writing
 \begin{equation}
 \label{A13}
   \Gamma_{Ri} = (\theta_{23}+\theta_{32}) \Gamma_{Ri} =
   \theta_{23} (\theta_{12} \Gamma_{Ri}) + 
   \theta_{32} (\theta_{13} \Gamma_{Ri})
 \end{equation} 
and subtracting the identities
 \begin{equation}
 \label{A14}
   \theta_{12} \theta_{23} \Gamma_{Ro} = 0 = 
   \theta_{13} \theta_{32} \Gamma_{R} \, .
 \end{equation}
This yields
 \begin{equation}
 \label{A15}
   \Gamma_{Ri} = \theta_{12} \theta_{32} \bigl(
   \Gamma_{122} + \Gamma_{211} - (\Gamma_{112} + \Gamma_{221}) \bigr)
              + \theta_{13} \theta_{32} \bigl(
   \Gamma_{122} + \Gamma_{211} - (\Gamma_{121} + \Gamma_{212}) \bigr)\, . 
 \end{equation}
Using Eqs.~(\ref{9c}) and (\ref{10}a,b) this is transformed into
 \begin{equation}
 \label{A16}
   \Gamma_{Ri} = \theta_{12} \theta_{23} 
                 \bigl( \tilde\Gamma_{112} + \tilde\Gamma_{221} 
                     - (\Gamma_{112} + \Gamma_{221}) \bigr)
               + \theta_{13} \theta_{32} 
                 \bigl(\tilde\Gamma_{121} + \tilde\Gamma_{212} 
                     - (\Gamma_{121} + \Gamma_{212}) \bigr) \, .
 \end{equation}

{\bf (iv)} We can summarize these results in coordinate space as follows:
 \begin{mathletters}
 \label{A17}
 \begin{eqnarray}
 \label{A17a}
   \Gamma_{R}  = \theta_{21} \theta_{13} \bar\rho_3
               + \theta_{23} \theta_{31} \bar\rho_1 \, ,
 \\
 \label{A17b}
   \Gamma_{Ri} = \theta_{12} \theta_{23} \bar\rho_3
               + \theta_{13} \theta_{32} \bar\rho_2 \, ,
 \\
 \label{A17e}
   \Gamma_{Ro} = \theta_{32} \theta_{21} \bar\rho_1
               + \theta_{31} \theta_{12} \bar\rho_2 \, ,
 \end{eqnarray}
 \end{mathletters}
where
 \begin{mathletters}
 \label{A18}
 \begin{eqnarray}
 \label{A18a}
   \bar\rho_1 &=& \tilde\Gamma_{122} + \tilde\Gamma_{211} 
           - (\Gamma_{122} + \Gamma_{211}) \, ,
 \\
 \label{A18b}
   \bar\rho_2 &=& \tilde\Gamma_{121} + \tilde\Gamma_{212} 
           - (\Gamma_{121} + \Gamma_{212}) \, ,
 \\
 \label{A18c}
   \bar\rho_3 &=& \tilde\Gamma_{112} + \tilde\Gamma_{221} 
           - (\Gamma_{112} + \Gamma_{221}) \, .
 \end{eqnarray}
 \end{mathletters}
{\bf (v)} In momentum space tilde conjugation reduces to complex conjugation
(see Eqs.~(\ref{6})), and the last three equations correspondingly
reduce to 
 \begin{mathletters}
 \label{A19}
 \begin{eqnarray}
 \label{A19a}
   \bar\rho_1 &=& - 2 i {\rm Im\, } (\Gamma_{122} + \Gamma_{211}) \, ,
 \\
 \label{A19b}
   \bar\rho_2 &=& - 2 i {\rm Im\, } (\Gamma_{121} + \Gamma_{212}) \, ,
 \\
 \label{A19c}
   \bar\rho_3 &=& - 2 i {\rm Im\, } (\Gamma_{112} + \Gamma_{221}) \, .
 \end{eqnarray}
 \end{mathletters}
Using the Fourier integral representation of the $\theta$ function
 \begin{equation}
 \label{A20}
   \theta_{ij} = -\frac {1}{2\pi i}\int_{-\infty}^{\infty}
   d\Omega \frac{e^{-i\Omega(t_i-t_j)}}{\Omega+i\epsilon} \,,
 \end{equation}
it is then straightforward to derive the following spectral integral 
representations in momentum space:
 \begin{mathletters}
 \label{A21}
 \begin{eqnarray}
 \label{A21a}
   \Gamma_{R}(\omega_1,\omega_2,\omega_3) &=&
   \frac{-i}{2\pi^2} \int_{-\infty}^{\infty}
   \frac{d\Omega_1 d\Omega_2}{\omega_2-\Omega_2+i\epsilon}
   \left(
   \frac{\rho_1(\Omega_1,\Omega_2,\Omega_3)}
        {\omega_1-\Omega_1-i\epsilon}
  +\frac{\rho_3(\Omega_1,\Omega_2,\Omega_3)}
        {\omega_3-\Omega_3-i\epsilon} \right) \, ,
 \\
 \label{A21b}
   \Gamma_{Ri}(\omega_1,\omega_2,\omega_3) &=&
   \frac{-i}{2\pi^2} \int_{-\infty}^{\infty}
   \frac{d\Omega_1 d\Omega_2}{\omega_1-\Omega_1+i\epsilon}
   \left(
   \frac{\rho_2(\Omega_1,\Omega_2,\Omega_3)}
        {\omega_2-\Omega_2-i\epsilon} 
  +\frac{\rho_3(\Omega_1,\Omega_2,\Omega_3)}
        {\omega_3-\Omega_3-i\epsilon} \right)\, ,
 \\
 \label{A21c}
   \Gamma_{Ro}(\omega_1,\omega_2,\omega_3) &=&
   \frac{-i}{2\pi^2}\int_{-\infty}^{\infty}
   \frac{d\Omega_1 d\Omega_2}{\omega_3-\Omega_3+i\epsilon}
   \left(
   \frac{\rho_1(\Omega_1,\Omega_2,\Omega_3)}
        {\omega_1-\Omega_1-i\epsilon} 
  +\frac{\rho_2(\Omega_1,\Omega_2,\Omega_3)}
        {\omega_2-\Omega_2-i\epsilon} \right)\, ,
 \end{eqnarray}
 \end{mathletters}
where $\omega_1+\omega_2+\omega_3 = \Omega_1 +\Omega_2 +\Omega_3 =0$, and
we used the new notation $\bar\rho_k = -2i \rho_k$ where (according to
(\ref{A19})) the $\rho_k$ are real functions. The spatial momenta on 
both sides of the equations are equal.

{\bf (vi)} In Ref.~\cite{CH96} it was shown that only two of the spectral 
densities $\rho_i$ are independent. This is consistent with the results 
derived here after realizing, by again using Eqs.~(\ref{9}) and (\ref{10}), 
that
 \begin{mathletters}
 \label{A22}
 \begin{eqnarray}
 \label{A22a}
   \theta_{12} \theta_{23} (\bar\rho_1 - \bar\rho_2 + \bar\rho_3) = 0 \, ,
 \\
 \label{A22b}
   \theta_{21} \theta_{13} (\bar\rho_2 - \bar\rho_1 + \bar\rho_3) = 0 \, ,
 \end{eqnarray}
 \end{mathletters}
such that one may substitute $\rho_3 = \rho_1 - \rho_2$ in $\Gamma_{R}$
and $\rho_3 = \rho_2 - \rho_1$ in $\Gamma_{Ri}$ (please note the opposite 
sign in the two cases!). This then yields Eqs.~(\ref{12}).

\subsection{1PI three-point functions}
\label{appa2}

The 1PI or truncated vertex functions $G_{abc}(k_1,k_2,k_3)$ are obtained
from the connected vertex functions $\Gamma_{abc}$ by truncating the
three external propagators:
 \begin{equation}
 \label{c0}
   G_{abc}(k_1,k_2,k_3) = {1\over i^3} D^{-1}_{aa'}(k_1)
   D^{-1}_{bb'}(k_2) D^{-1}_{cc'}(k_3) \Gamma_{a'b'c'}(k_1,k_2,k_3)\, .
 \end{equation}  
They satisfy the identity 
 \begin{equation}
   \sum_{a,b,c=1}^2 G_{abc}=0 \, ,
 \label{c2}
 \end{equation}
and the three retarded 1PI vertices are given by \cite{chou,CH96}
 \begin{mathletters}
 \label{c1}
 \begin{eqnarray}
 \label{c1a}
   G_{R} &=& G_{111}+G_{112}+G_{211}+G_{212} \, ,
 \\
 \label{c1b}
   G_{Ri} &=& G_{111}+G_{112}+G_{121}+G_{122} \, ,
 \\
 \label{c1c}
   G_{Ro} &=& G_{111}+G_{121}+G_{211}+G_{221} \, .
 \end{eqnarray}
 \end{mathletters}
These relations differ from (\ref{5}) and (\ref{11}) only by sign
factors $(-1)^{a+b+c-3}$. In momentum space we have instead of
(\ref{6}) \cite{kob2}
 \begin{mathletters}
 \label{c3}
 \begin{eqnarray}
 \label{c3a}
   \tilde G_{111}(k_1,k_2,k_3) &=-& G_{111}^*(k_1,k_2,k_3)
   = G_{222} (k_1,k_2,k_3) \, ,
 \\
 \label{c3b}
   \tilde G_{121}(k_1,k_2,k_3) &=& -G_{121}^*(k_1,k_2,k_3)
   = e^{\beta \omega_2}\,G_{212}(k_1,k_2,k_3) \, ,
 \\
 \label{c3c}
   \tilde G_{211}(k_1,k_2,k_3) &=&- G_{211}^*(k_1,k_2,k_3)
   = e^{\beta \omega_1}\,G_{122}(k_1,k_2,k_3) \, ,
 \\
 \label{c3d}
   \tilde G_{112}(k_1,k_2,k_3) &=& -G_{112}^*(k_1,k_2,k_3) 
   = e^{\beta\omega_3}\, G_{221}(k_1,k_2,k_3)
 \, ,
 \end{eqnarray}
 \end{mathletters} 
where ``tilde conjugation'' is defined in the same way as for the
connected vertex. 

From identities like $\theta_{32} \theta_{21} (G_R+G_{Ri})=0$
involving conflicting $\theta$-functions one can derive
largest and smallest time equations for the 1PI vertices similar to
those derived in Sec.~\ref{sec2b} for the connected vertices.
One finds that the sign factor $(-1)^{a+b+c-3}$ simply carries over,
changing all relative minus signs in Eqs.~(\ref{8})-(\ref{10}) into
plus signs. For example, Eq.~(\ref{8}) turns into 
 \begin{equation}
 \label{8'}
   \theta_{32}\theta_{21}(G_{111}+G_{112})=0\, .
 \end{equation}

By following the same procedure as for the connected functions in
Appendix~\ref{appa1} one obtains for the truncated functions the
following relations in coordinate space:
 \begin{mathletters}
 \label{c17}
 \begin{eqnarray}
 \label{C17a}
   G_{R}  = \theta_{21} \theta_{13} \bar\rho_3'
               + \theta_{23} \theta_{31} \bar\rho_1' \, ,
 \\
 \label{C17b}
   G_{Ri} = \theta_{12} \theta_{23} \bar\rho_3'
               + \theta_{13} \theta_{32} \bar\rho_2' \, ,
 \\
 \label{C17e}
   G_{Ro} = \theta_{32} \theta_{21} \bar\rho'_1
               + \theta_{31} \theta_{12} \bar\rho'_2 \, ,
 \end{eqnarray}
 \end{mathletters}
where
 \begin{mathletters}
 \label{C18}
 \begin{eqnarray}
 \label{c18a}
   \bar\rho'_1 &=&\tilde G_{211} -\tilde G_{122}   
           + G_{211} -G_{122}  \, ,
 \\
 \label{c18b}
   \bar\rho'_2 &=& \tilde G_{121} - \tilde G_{212} 
          + G_{121} -G_{212} \, ,
 \\
 \label{c18c}
   \rho'_3 &=& \tilde G_{112} - \tilde G_{221} 
           + G_{112} - G_{221} \, .
 \end{eqnarray}
 \end{mathletters}
In momentum space, by making use of Eqs.~(\ref{c3}), the last three
equations reduce to 
 \begin{mathletters}
 \label{C19}
 \begin{eqnarray}
 \label{C19a}
   \bar\rho'_1 &=& - 2 i \, {\rm Im\, } (G_{122} - G_{211}) \, ,
 \\
 \label{C19b}
   \bar\rho'_2 &=& - 2 i \, {\rm Im\, } (G_{212} - G_{212}) \, ,
 \\
 \label{C19c}
   \bar\rho'_3 &=& - 2 i \, {\rm Im\, } (G_{221} -G_{221}) \, .
 \end{eqnarray}
 \end{mathletters}
Using further Eq.~(\ref{c2}), the largest and smallest time equations,
and Eqs.~(\ref{c3}) for the truncated functions one shows that
 \begin{mathletters}
 \label{c22}
 \begin{eqnarray}
 \label{c22a}
   \theta_{12} \theta_{23} (\bar\rho'_1 -\bar \rho'_2 +\bar \rho'_3) = 0 \, ,
 \\
 \label{c22b}
   \theta_{21} \theta_{13} (\bar\rho'_2 - \bar\rho'_1 +\bar \rho'_3) = 0 \, .
 \end{eqnarray}
 \end{mathletters}
Inserting these into Eqs.~(\ref{c17}) and transforming to momentum
space one obtains the spectral integrals for the truncated vertex
functions given in Eqs.~(\ref{c21}). Please note that, up to the
different definition of the spectral densities, they are formally
identical with the spectral representations (\ref{12}) for the
corresponding connected vertex functions.

\section{Symmetries of the three-point spectral densities}
\label{appb}

In addition to Eq.~(\ref{25a}), which holds only for 3-point vertices
with three identical external legs, there are some useful other
symmetries for the 3-point spectral densities. Inserting the
propagators (\ref{2}) into (\ref{16}) and using twice the relation
(\ref{reldis}) one obtains, without any further manipulations, the
expression
 \begin{eqnarray}
 \label{b1}
   \rho'_1(p,q) &=& g^3 \int {d^n s\over (2\pi)^{n-3}} 
          {\rm sgn}(s_0)\,{\rm sgn}(s_0+q_0)\, {\rm sgn}(s_0-p_0)\, 
 \nonumber\\
   && \quad \times \,
      \delta(s^2-m^2)\, \delta\bigl( (s+q)^2 - m^2 \bigr)\,
      \delta\bigl( (s-p)^2 - m^2 \bigr)\,
 \nonumber\\
   && \quad \times \,
      \Bigl[{\textstyle{1\over 2}}\Bigl(n(s_0)-n(s_0-p_0)\Bigr) 
      + n(q_0) \Bigl( n(s_0) - n(s_0+q_0) \Bigr) 
 \nonumber\\
   && \quad \ \ \ + n(p_0+q_0) \Bigl( n(s_0+q_0) - n(s_0-p_0) \Bigr)
   \Bigr]\, .
 \end{eqnarray}
By reversing the sign of the spatial integration variable, $\bbox{s}
\mapsto -\bbox{s}$, one immediately reads off 
 \begin{equation}
 \label{b2}
  \rho'_1(p_0,\bbox{p};q_0,\bbox{q}) = 
  \rho'_1(p_0,-\bbox{p};q_0,-\bbox{q})\, .
 \end{equation}
With the additional help of the identity $n(-x) = - \bigl( 1+n(x)
\bigr)$ one shows similarly that 
 \begin{equation}
 \label{b3}
  \rho'_1(-p,-q) = - \rho'_1(p,q)\, .
 \end{equation}
The combination of these two identities gives
 \begin{equation}
 \label{b4}
  \rho'_1(-p_0,\bbox{p};-q_0,\bbox{q}) = 
  - \rho'_1(p_0,\bbox{p};q_0,\bbox{q})\, ,
 \end{equation}
i.e. the spectral density is odd under simultaneous sign change of
both frequencies. This generalizes a similar condition for the
spectral density for the single-particle propagator. The above
symmetry relations are consistent with Eqs.~(4) and (5) in Ref.~\cite{Aur}.


\newpage

\begin{thebibliography}{99}

\bibitem[*]{email}
  On leave of absence from Hua-Zhong Normal University, Wuhan, China;
  e-mail address: {\tt defu.hou@rphs1.physik.uni-regensburg.de}
\bibitem{LvW87} 
  N.P. Landsman and Ch.G. van Weert, Phys. Rep. {\bf 145}, 141 (1987). 
\bibitem{K89} 
  J.I. Kapusta, Finite Temperature Field Theory, Cambridge University
  Press, 1989.
\bibitem{HST84} 
  A.Hosoya, M. Sakagami, and M. Takao, Ann. Phys. (N.Y.) {\bf 154},
  229 (1984), and references therein.
\bibitem{J93}
  S. Jeon, Phys. Rev. D{\bf 47}, 4586 (1993).
\bibitem{WHZ96} 
  E. Wang, U. Heinz, and X.F. Zhang, Phys. Rev. D{\bf 53}, 5978 (1996).
\bibitem{keld} 
  L.V. Keldysh, Zh. Eksp. Teor. Fiz. {\bf 47}, 1515 (1964) [JETP {\bf 20},
  1018 (1965)].
\bibitem{chou}
  K.-C. Chou, Z.-B. Su, B.-L. Hao and L. Yu, Phys. Rep. {\bf 118}, 1 (1985).
\bibitem{henning} 
  P.A. Henning, Phys. Rep. {\bf 253}, 235 (1995). 
\bibitem{pisarski}
  E. Braaten, R.D. Pisarski, Nucl. Phys. B{\bf 337}, 569 (1990); 
  and B{\bf 339}, 310 (1990). 
\bibitem{altherr}
 T. Altherr, T. Grandou and R.D. Pisarski, Phys. Lett. B{\bf 271}, 183 (1991).
\bibitem{taylor}
  J.C. Taylor, Phys. Rev. D{\bf 48}, 958 (1993).
\bibitem{evan1}
  T.S. Evans, Phys. Lett. B{\bf 249}, 286 (1990).
\bibitem{kob}
  R. Kobes and G.W. Semenoff, Nucl.Phys. B{\bf 260}, 714 (1985);
  and B{\bf 272}, 329 (1986). 
\bibitem{kob1}
  R. Kobes, Phys. Rev. D{\bf 43}, 1269 (1991).
\bibitem{BDN96} 
  P.F. Bedaque, A. Das, and S. Naik, Los Alamos eprint archive hep-ph/9603325.
\bibitem{Gelis}
  F. Gelis, Los Alamos eprint archive hep-ph/9701410.
\bibitem{L96} 
  P.V. Landshoff, Cambridge preprint DAMTP 96/64.
\bibitem{CH96} 
  M.E. Carrington and U. Heinz, Los Alamos eprint archive
  hep-th/9606055, Z. Phys. C, in press.
\bibitem{evan2}
  T.S. Evans, Phys. Lett. B{\bf 252}, 108 (1990); and Nucl. Phys. 
  B{\bf 374}, 340 (1992).
\bibitem{kob2}
  R. Kobes, Phys. Rev. D{\bf 42}, 562 (1990).
\bibitem{Umez}
  H. Umezawa, H. Matsumoto, and M. Takichi, {\it Thermo Field Dynamics
  and Condensed States} (North Holland, Amsterdam, 1982).
\bibitem{bair}
  R. Baier, B. Pire, and D. Schiff, Z. Phys. C{\bf 51}, 581 (1991).
\bibitem{ashida}
  N. Ashida, A. Ni\'egawa, H. Nakkagawa, and H. Yokota,
  Phys. Rev. D{\bf 44}, 473 (1991); and {\bf 45}, 1432(E) (1991).
\bibitem{Aur}
    P. Aurenche, E. Petitgirard and T.R. Gaztelurrutia,
    Phys. Lett. B{\bf 297}, 337 (1992).
\bibitem{mrow}
    S. Mr\'owczy\'nski, Los Alamos eprint archive hep-th/9702022,
    Phys. Rev. D (15 Aug. 1997), in press.
\bibitem{schulz} 
    U. Kraemmer, A.K. Rebhan, and H. Schulz, Ann. Phys. (N.Y.) {\bf 238},
    286 (1995).
\bibitem{BP92}
    E. Braaten and R.D. Pisarski, Phys. Rev. D{\bf 45}, R1827 (1992).
\bibitem{FT92}
    J. Frenkel and J.C. Taylor, Nucl. Phys. B{\bf 374}, 156 (1992).
\bibitem{leupold}
    We would like to thank S. Leupold for an enlightening discussion
    on this point.

\end{thebibliography}
\end{document}